\documentclass[11pt]{article}
\usepackage{amsmath,amssymb}
\usepackage{graphicx,color}
\usepackage{cite}

\newcommand{\corrected}[1]{{#1}}

\def\cC{{\mathcal C}}

\def\cK{{\mathcal K}}
\def\cL{{\mathcal L}}
\def\cN{{\mathcal N}}
\def\cR{{\mathcal R}}

\def\({\left(}
\def\){\right)}

\newcommand{\pd}{\partial}

\newcommand{\de}{\partial}
\newcommand{\be}{\begin{equation}}
\newcommand{\ba}{\begin{eqnarray}}
\newcommand{\ea}{\end{eqnarray}}
\newcommand{\ee}{\end{equation}}

\newcommand{\f}{\frac}
\newcommand{\s}{\sqrt}

\newcommand{\ap}{\alpha}

\newcommand{\ddd}{\cdot\cdot\cdot}
\newcommand{\no}{\nonumber \\}

 \def\al{\alpha'}
 \def\de{\partial}

 \def\f {\frac}
 
 \def\ap{\alpha}

 \def\ddd{\cdot\cdot\cdot}
 \def\no{\nonumber \\}

\newcommand{\sfrac}[2]{{{#1}/{#2}}}

\begin{document}

\begin{titlepage}
\thispagestyle{empty}

\begin{flushright}
IPMU11-0119
\end{flushright}

\begin{center}
\noindent{\large \textbf{
Higher Derivative Corrections to Holographic Entanglement Entropy \\
for AdS Solitons}}\\
\vspace{2cm} \noindent{Noriaki
Ogawa\footnote{e-mail: noriaki.ogawa@ipmu.jp}
and Tadashi Takayanagi\footnote{e-mail: tadashi.takayanagi@ipmu.jp}}\\
\vspace{1cm} {\it
Institute for the Physics and Mathematics of the Universe (IPMU), \\
 University of Tokyo, Kashiwa, Chiba 277-8582, Japan\\
} \vspace{2cm}
\end{center}

\begin{abstract}
We investigate the behaviors of holographic entanglement entropy
for AdS soliton geometries in the presence of higher derivative corrections.
We calculate the leading higher derivative corrections for the AdS$_5$ setup in type IIB string
and for the AdS$_{4,7}$ ones in M-theory. We also study the holographic entanglement entropy
in Gauss-Bonnet gravity and study how the confinement/deconfinement phase transition observed in AdS solitons is affected by the higher derivative corrections.
\end{abstract}

\end{titlepage}

%\newpage

\tableofcontents

\section{Introduction}

The entanglement entropy
is a useful universal quantity when we
would like to employ the AdS/CFT correspondence \cite{Maldacena} to
study non-perturbative aspects of string theory as quantum gravity.
In \cite{RT}, it has been proposed that the entanglement entropy in
conformal field theories (CFTs) can be calculated from the area of
minimal surface in anti de-Sitter (AdS) spaces. It is explicitly
given by the formula \be S_A=\f{\mbox{Area}(\gamma_A)}{4G_N},
\label{area} \ee where $G_N$ is the Newton constant in the Einstein
gravity on the AdS space; $S_A$ is the entanglement entropy for the
subsystem $A$, which can be chosen arbitrarily; $\gamma_A$ is the
codimension two minimal area surface which ends on the boundary of
$A$. Also we require $\gamma_A$ is homologous to $A$. Even though
there has been no precise proof of this formula
(\ref{area}), many non-trivial evidences have been accumulated until
now. For example, the strong subadditivity, which is one of the most
important inequality satisfied by the entanglement entropy, has been shown in
\cite{HeTa}. Quite recently, the paper \cite{HHM}  showed that
all known inequalities are satisfied by the holographic entanglement entropy (HEE)
given by (\ref{area}) and moreover HEE leads to more constrained inequalities than
entanglement entropies in general quantum field theories. Also, the agreements
of the logarithmic terms in both side
has been shown in \cite{RT,Sol,LNST,CaHu,Dowker,Solog,CHM}. Moreover,
in \cite{He}, a highly non-trivial consistency check has been made
when the subsystem $A$ is two intervals \cite{CCT} in two dimensional CFTs by employing
so called Renyi entropies. Refer to
\cite{NRT} and references therein for more evidences. For quantum field theoretic
analysis of entanglement entropy refer e.g. to the excellent reviews \cite{ReCH,ReCC}.

The holographic entanglement entropy (\ref{area}) assumes that the
gravity on the AdS space is defined by the Einstein gravity with a
negative cosmological constant. In the CFT side, this means that we
are taking the large 't Hooft coupling limit. To analyze a full
quantum gravity which appears in string theory and to extend the
results to the weak coupling region of the CFT dual, we need to include
quantum corrections which are effectively described by higher
derivative terms. In the recent papers \cite{HMM,deBoer:2011wk}
studied the corrections of holographic entanglement entropy in the
presence of higher derivative terms (see also \cite{MySi} for
related discussions). Though it is very complicated to find a holographic
entanglement entropy formula for general higher derivative
corrections as noted in \cite{HMM}, a class of theories called Lovelock
gravities, the authors of \cite{HMM,deBoer:2011wk} found important evidences that
natural extensions of (\ref{area}) give the correct holographic
entanglement entropy. In the $\cR^2$ Gauss-Bonnet gravity, which is
the simplest example of Lovelock gravities, it is given by the
following formula (this expression itself has been already
speculated in \cite{Fu})
 \be S_A=\f{1}{4G_N}\int_{\gamma_A}\s{h}(1+\eta L^2 \cR),
\label{hheein} \ee where $\eta$ is the coupling constant of
Gauss-Bonnet term, which will be explained later; $h$ is the induced
metric on the three dimensional surface $\gamma_A$ chosen
arbitrarily; $\cR$ is the {\it intrinsic} curvature of $h$. We again
require that $\gamma_A$ has a boundary which coincides with that of
$A$ and we need to add a boundary Gibbons-Hawking term to
(\ref{hheein}) in actual calculations. The profile of $\gamma_A$ is
determined by minimizing the functional (\ref{hheein}).

  The purpose of this paper is to investigate further the properties of higher
  derivative corrections of the entanglement entropy. In the first half of this paper,
  we will consider explicit setups of string/M-theory and calculate the
higher derivative corrections of the holographic entanglement
entropy. Since the leading higher derivative correction is the
quartic polynomial of the Weyl curvature \cite{HDS,HDM}, this is not
in the class of Lovelock gravities. To obtain non-trivial results in
a tractable way, we will concentrate on the examples of AdS soliton
spaces, which are dual to confining gauge theories \cite{Wi}, and
choose the subsystem as just the half of the total space.
In the type IIB string case, this is dual to the subleading
corrections to the entanglement entropy in the $\cN=4$ super
Yang-Mills (SYM) on a circle with the anti-periodic boundary
condition for fermions. In the latter half of this paper,
we will study the holographic entanglement entropy
in the 4D and 5D Gauss-Bonnet gravity, especially when the subsystem
$A$ is defined by a strip. We will analyze the behavior of the
entropy when we change the width of the strip and see how the phase transition is
affected by the higher derivatives. The holographic entanglement entropy for AdS solitons
without higher derivative corrections has been already studied in \cite{NiTa,KKM,PP}.
Later, qualitatively similar results has been recently reproduced in lattice gauge theory approaches \cite{La,Lb,Lc}.

This paper is organized as follows: In section \ref{sec:string} we calculate the
higher derivative corrections of holographic entanglement entropy
for the AdS$_5$ soliton in IIB string and for the AdS$_{4,7}$ soliton in
M-theory. In section \ref{sec:gb}, we compute the holographic
entanglement entropy for AdS soliton in the Gauss-Bonnet gravity in
four and five dimension. We choose the subsystem $A$
to be a half space in all of the above calculations. In section \ref{sec:gbpt},
we analyze the holographic entanglement entropy for AdS
soliton in the five dimensional Gauss-Bonnet gravity when $A$ is
give by a strip. In section \ref{sec:summary}, we summarize our conclusions.

\section{Holographic Entanglement Entropy with Higher Derivative Corrections
in String and M Theories}
\label{sec:string}

In this section, we will investigate the leading higher derivative
corrections to the holographic entanglement entropy in string theory and M-theory.
We consider AdS$_5$ soliton geometry in type IIB superstring and
AdS$_4$ and AdS$_7$ soliton geometries in M-theory. We choose the subsystem A as the
half of the total space.

\subsection{Type IIB Superstring}

Consider the AdS$_5\times $S$^5$ solution in type IIB string theory.
The relevant part of the type IIB supergravity action with the
leading $\al$ correction looks like in the Euclidean signature
\cite{HDS} \be I=-\f{1}{16\pi G^{(10)}_N}\int
dx^{10}\s{g}\left[\cR-\f{1}{2}(\de_\mu\phi)^2-\f{1}{4\cdot 5\!}
F_5^2+ \dotsc +
\gamma\cdot e^{-\f{3}{2}\phi}W + \dotsc \right], \label{IIBaction} \ee
where we keep the relevant
terms which are linear with respect to $\gamma$, which is supposed
to be very small.
There are also other such terms depending on other antisymmetric fields,
represented by dots, but they do not have any contributions to our discussions below.
$W$ is defined by the following $\cR^4$ term by using the Weyl
curvature $\cC$ \be
W=\cC^{hmnk}\cC_{pmnq}\cC_h^{\ rsp}\cC^q_{\ rsk}+\f{1}{2}\cC^{hkmn}\cC_{pqmn}\cC_h^{\ rsp}\cC^q_{\ rsk},
\label{Wfunc} \ee and the constant $\gamma$ is give by \be
\gamma=\f{1}{8}\zeta(3) \al^3. \ee

The AdS$_5\times$ S$^5$ is the solution to this theory with the
higher derivatives. This type IIB string background is dual to the
four dimensional $\cN=4$ $SU(N)$ super Yang-Mills theory
\cite{Maldacena}. The standard dictionary tells us that the AdS
radius $L$ is related to the 't Hooft coupling $\lambda=2\pi g_s N$
by \be L^4=4\pi g_s\al^2 N=2\lambda \al^2. \ee Therefore taking into
account the linear order of $\gamma$ is to consider the next leading
order correction in the strong coupling expansion of $\lambda$. The
Yang-Mills coupling is related to the string coupling \be
g^2_{YM}=2\pi g_s. \ee The ten dimensional Newton constant is given
by \be G^{(10)}_N=8\pi^6\al^4g_s^2=\f{\pi^4L^8}{2N^2}. \ee

\subsubsection{AdS$_5$ soliton and
the large 't Hooft coupling limit of $\mathcal{N}=4$ SYM}

Consider the AdS$_5$ soliton background, which is dual to the
$\cN=4$ SYM compactified on a circle $S^1$ (radius $R$) with the
anti-periodic boundary condition for fermions \cite{Wi}. Its leading
order solution (i.e. $\gamma=0$) is given by \be
ds^2=\f{r^2}{L^2}[dt_E^2+dx^2+dy^2]+\f{L^2}{r^2}\f{dr^2}{1-r^4_0/r^4}
+\f{r^2}{L^2}(1-r_0^4/r^4)d\theta^2+L^2d\Omega^2_5. \ee The
periodicity of $\theta$ is given by $\theta\sim\theta +2\pi R$ with
\be R=\f{L^2}{2r_0}. \ee

We define the subsystem $A$ by the following half space
on the time slice $t_E=0$ \be x>0,\ \ \  -\infty<y<\infty,\ \ \
0\leq \theta<2\pi R. \ee The contribution of holographic
entanglement entropy in the leading order of $\gamma$ is given by
the area law formula (\ref{area}). In the AdS soliton case, we
obtain explicitly \cite{NiTa} \ba
S^{(0)}_A&=&\f{\mbox{Area}(\gamma_A^{(5)})}{4G^{(5)}_N}=\f{\pi^3L^4}{8G^{10}_N}\f{\pi
L^2}{r_0}(r^2_\infty-r^2_0)V_1,\no &=&
\f{N^2}{4L^2}\cdot\f{r^2_\infty}{r_0}V_1-\f{N^2V_1}{8R},
\label{trsol} \ea
where we employed
\be
G^{(5)}_N=\f{G^{(10)}_N}{\mbox{vol}(S^5)}=\f{\pi L^3}{2N^2},
\ee
using $\mbox{vol}(S^5)=\pi^3L^5$.

\subsubsection{Higher derivative corrections to the entanglement entropy}

Next we would like to calculate the higher derivative corrections to
the previous calculation (\ref{trsol}). To calculate the holographic
entanglement entropy, we place a deficit angle $\delta=2\pi(1-n)$ on
the three dimensional surface $\gamma_A$ defined by $x=t_E=0$
\cite{CaCa,Fu,RT}. When the extrinsic curvature of the submanifold
$\gamma_A$ is vanishing, the curvature tensor behaves like (assuming
$1-n$ is infinitesimally small) \cite{SoFu}
\ba &&
\cR_{\mu\nu\rho\sigma}=\cR^{(0)}_{\mu\nu\rho\sigma}
+2\pi(1-n)(N_{\mu\rho}N_{\nu\sigma}-N_{\mu\sigma}N_{\nu\rho})\delta(\gamma_A),
\no && \cR_{\mu\nu}=\cR^{(0)}_{\mu\nu}
+2\pi(1-n)N_{\mu\nu}\delta(\gamma_A), \no &&
\cR=\cR^{(0)}+4\pi(1-n)\delta(\gamma_A), \label{curv}
\ea
where we defined $N_{\mu\nu}=n^{t_E}_\mu n^{t_E}_\nu+n^{x}_\mu n^{x}_\nu$.
The unit vectors $n^{t_E}_\mu$ and $n^{x}_\mu$ are orthogonal to the
surface $\gamma_A$. $\cR^{(0)}_{\mu\nu}$ etc. denotes
 the original curvature tensor without the deficit angle.

By plugging (\ref{curv}) into (\ref{Wfunc}), we obtain%
\footnote{
By naive substitution, we have several terms including
$\delta(\gamma_A)^k$ $(k\ge 2)$ in the subleading part $O((1-n)^2)$.
Precisely speaking, we should regularize the $\delta$-functions at first
and take the limit in the last stage of the calculation.
But anyway this does not affect our results below.
}
\be
W=\f{180 r_0^{16}}{L^8r^{16}}
+ 2\pi(n-1)\delta(\gamma_A)\cdot\f{40r^{12}_0}{L^6r^{12}}
+ O(\gamma) + O\((1-n)^2\).
\ee
Therefore, the contribution from the $W$-term is estimated to be
\ba
S^{(W)}_A&=&-\f{\de}{\de n}\log\left.\f{Z_n^{(W)}}{\(Z_1^{(W)}\)^n}\right|_{n=1}\no
&=&-\f{\gamma}{16\pi G^{(10)}_N}\int_{\gamma_A} dx^8\s{g_{\gamma_A}}\left[\f{80\pi
r^{12}_0}{L^6r^{12}}\right]\no
&=&-\f{5\gamma r^{12}_0}{L^7G^{(5)}_N}
(2\pi RV_1)\int^{r_{\infty}}_{r_0}\f{dr}{r^{11}}\no
&=&
-\frac{\gamma N^2V_1}{2RL^6},
\label{wterm}
\ea
where
$\s{g_{\gamma_A}}=rL^4d\Omega_5$
is the induced volume factor on $\gamma_A$.

Next we need to consider the higher derivative corrections of the
metric. It is given by \be
ds^2=e^{-\f{10}{3}\nu}ds^2_{5}+e^{2\nu}(d\Omega_5)^2, \ee where
$\nu$ is order $O(\gamma)$. The five dimensional part is given by a
modification of the AdS soliton metric \cite{GKP,PaTh} \be
ds^2=f(r)d\theta^2+\f{1}{g(r)}dr^2+r^2(dt^2_E+dx^2+dy^2), \ee where
\ba && f(r)=\f{r^4-r^4_0}{r^2}\left(1+\gamma\cdot\ap(r)\right),\no
&& g(r)=\f{r^4-r^4_0}{r^2}\left(1+\gamma\cdot\beta(r)\right). \ea
The functions $\ap(r)$ and $\beta(r)$ are defined by \ba &&
\ap(r)=-\f{15}{L^6}\left(5\f{r^4_0}{r^4}+5\f{r^8_0}{r^8}-3\f{r^{12}_0}{r^{12}}\right),\no
&&
\beta(r)=-\f{15}{L^6}\left(5\f{r^4_0}{r^4}+5\f{r^8_0}{r^8}-19\f{r^{12}_0}{r^{12}}\right).
\ea

The radius in the $\theta$-direction can be found by requiring the
smoothness of the metric
\be
2\pi R=\f{\pi L^2}{r_0}-\f{15\pi \gamma}{r_0 L^4}+O(\gamma^2).
\ee

Since we already took care of the $W$ term in (\ref{wterm}), here we
only consider the contribution from the Einstein-Hilbert term. This
is simply given by the area law formula as \ba
S^{(EH)}_A&=&\f{V_1}{4G_N^{(5)}}\cdot\f{2\pi R}{L}\cdot
\left(\f{r^2_\infty}{2}-\f{r^2_0}{2}-12\gamma\f{r^2_0}{L^6}\right)\no
&=& \f{V_1 N^2R}{2L^4}r^2_\infty-\f{N^2V_1}{8R}+\f{3\gamma N^2V_1}{4RL^6}
+ O(\gamma^2).
\ea

Finally the total expression of $S_A$ up to $O(\gamma)$ is given by
\ba S_A=S^{(EH)}_A+S^{(W)}_A&=&\f{V_1 N^2R}{2L^4}r^2_\infty
-\f{N^2V_1}{8R}+\f{\gamma N^2V_1}{4RL^6}\no &=&
(\mbox{div.})+\f{N^2V_1}{R}\left(-\f{1}{8}
+\f{\zeta(3)}{32}(2\lambda)^{-3/2}\right), \ea
where div. denotes the quadratically divergent term.
We are interested in its finite term, which is independent of the UV
cut off.
This final result means that the
entropy increases as the 't Hooft coupling gets decreased. Indeed,
the free field theory result obtained in \cite{NiTa} is given by \ba
S^{free}_A=(\mbox{div.})-\f{N^2V_1}{12R},\ea and matches with this
expectation from our gravity calculation.

\subsection{M-Theory}

It will be interesting to compute the higher derivative corrections
of the holographic entanglement entropy in M-theory backgrounds.
The one-loop corrected action of the 11D supergravity is given by
\cite{HDM}
\be I=-\f{1}{2\kappa_{11}^2}\int
dx^{11}\s{g}\left[\cR+\dotsc
+\gamma\cdot W + \dotsc
\right],
\label{Maction} \ee
where the dots represent the terms depending on antisymmetric forms again.
$W$ is given by the same form as \eqref{Wfunc}, and $\gamma$ is
\be
\gamma = \f{4\pi^2}{3}\kappa_{11}^{\f{4}{3}}.
\label{Mgamma}
\ee
The 11D Newtonian constant $G_N^{(11)}$
and the Planck length $\ell_P$
are related to $\kappa_{11}$ as
\be
8\pi G_N^{(11)} = \kappa_{11}^2 = 2^4\pi^5\ell_P^9.
\ee

\subsubsection{M2-brane Soliton}
Consider the AdS$_4$ soliton which is dual to a three dimensional CFT
on $N$ M2-branes compactified on a circle.
First of all, upon dimensional reduction on an $S^7$,
the correction term $I^{(W)}$ in \eqref{Maction} is reduced to%
\footnote{
Here we assume that the 11D one-loop corrected action \eqref{Maction}
may be dealt with classically on the spherical compactified background
and on its dimensional reduction. We believe this is plausible
because a sphere does not have noncontractable cycles,
which would bring about new loop diagrams.
The authors thank Shinji Hirano and Masaki Shigemori for discussions on this issue.
}
\be
I^{(W)} = -\frac{1}{16\pi G_N^{(4)}}\int d^4x\sqrt{g_4}\,\gamma W_4,
\ee
where $W_4$ is given by the same form as \eqref{Wfunc} in 4D
and the 4D Newtonian constant $G_N^{(4)}$ is given by
\be
G_N^{(4)} = \f{G_N^{(11)}}{\mbox{vol}(S^7)}.
\ee

Under this dimensional reduction,
the one-loop corrected metric
of the AdS$_4$ soliton geometry reads \cite{CaKl} \be
ds^2_4=f(r)d\theta^2+\f{dr^2}{g(r)}+\f{r^2}{L^2}(dt^2_E+dx^2), \ee
where \ba && f(r)=\f{r^2}{L^2}\left(1-\f{r^3_0}{r^3}\right)\cdot
(1+\gamma\cdot \ap(r)),\no &&
g(r)=\f{r^2}{L^2}\left(1-\f{r^3_0}{r^3}\right)\cdot (1+\gamma\cdot
\beta(r)),\no &&
\ap(r)=-\f{1}{4L^6}\left(17\f{r^3_0}{r^3}+17\f{r^6_0}{r^6}-11\f{r^9_0}{r^9}\right),\no
&&
\beta(r)=-\f{1}{4L^6}\left(17\f{r^3_0}{r^3}+17\f{r^6_0}{r^6}-67\f{r^9_0}{r^9}\right).
\ea The AdS radius $L$ is related to the number $N$ of M2-branes by
\be L^9=N^{3/2}\f{\kappa^2_{11}}{2^{17/2}\cdot\pi^5}, \label{M2L}\ee
and the
volume of $S^7$ is given by \be \mbox{Vol}(S^7)=\f{\pi^4}{3}(2L)^7.
\ee The radius in the $\theta$-direction is found to be \be 2\pi
R=\f{4\pi L^2}{3r_0}-\f{5\pi\gamma}{3L^4r_0}. \ee
Using \eqref{Mgamma} \eqref{M2L}, the expansion parameter $\gamma$ is reexpressed as
\be
\gamma=\f{2^{23/3}\cdot\pi^{16/3}}{3}\cdot \f{L^6}{N}.
\ee
It is useful to note
\be
\f{\mbox{Vol}(S^7)}{\kappa^2_{11}}=\f{1}{2^{3/2}\cdot 3\pi}\cdot\f{N^{3/2}}{L^2}.
\ee

In the presence of the deficit angle $\delta=2\pi(1-n)$, the higher derivative term $W_4$ behaves like
\be
W_4=
\f{9}{2}\f{r_0^{12}}{L^8r^{12}}
+2\pi(1-n)\delta(\gamma_A)\cdot\f{6r_0^9}{L^6r^9}
+O(\gamma) + O\((1-n)^2\),
\ee
using the uncorrected AdS$_4$ soliton metric.
This leads to the contribution
\be
S^{(W)}_A=\f{3\pi^2 \gamma R r_0}{2\kappa^2_{11}L^6}\cdot\mbox{Vol}(S^7).
\ee
The contribution from the Einstein-Hilbert action can be found by applying the area formula to the corrected metric
\ba
S^{(EH)}_A &=& \f{4\pi^2 R}{\kappa^2_{11}}\cdot\mbox{Vol}(S^7)\cdot \int^{r_{\infty}}_{r_0}dr \left(1-\f{7\gamma}{L^6}\f{r^9_0}{r^9}\right) \no
&=&\f{4\pi^2 R}{\kappa^2_{11}}\cdot\mbox{Vol}(S^7)\cdot
\(r_{\infty} - \(1 + \f{7\gamma}{8L^6}\)r_0\)
\ea
Finally the total contribution to $S_A$ is
\ba
S^{(EH)}+S^{(W)}
&=&\f{N^{3/2}}{L^2}\f{1}{2^{3/2}\cdot 3\pi}\left(4\pi^2 Rr_{\infty}-2\pi^2 Rr_0\(2+\f{\gamma}{L^6}\)\right) \no
&=&\f{N^{3/2}}{L^2}\f{1}{2^{3/2}\cdot 3\pi}\left(4\pi^2 Rr_{\infty}
- \f{8\pi^2L^2}{3} + \f{2\pi^2\gamma}{L^4} + O(\gamma{})^2\right).
\ea
In the end, the $1/N$ correction to the finite term is found to be
\ba
\left(S^{(EH)}+S^{(W)}\right)_{finite}&=&\f{N^{3/2}}{L^2}\f{1}{2^{3/2}\cdot 3\pi}
\left(-\f{8\pi^2L^2}{3}+2\pi^2\f{\gamma}{L^4} + O(\gamma^2)\right)\no
&=&-\f{2^{3/2}\pi}{9}N^{3/2}+\f{\pi^{19/3}2^{43/6}}{9}N^{1/2} + O(N^{-1/2}).
\ea

\subsubsection{M5-brane Soliton}

Finally we would like to study the AdS$_7$ soliton dual to the six
dimensional SCFT of $N$ M5-branes on a supersymmetry breaking circle.
In a similar way to the AdS$_4$ case,
we reduce the 11D theory on $S^4$ and then
the one-loop corrected 7D metric reads \cite{CaKl} \be
ds^2_7=f(r)d\theta^2+\f{dr^2}{g(r)}+\f{r^2}{L^2}(dt^2_E+\sum_{i=1}^{4}dx_i^2),
\ee where \ba && f(r)=\f{r^2}{L^2}\left(1-\f{r^6_0}{r^6}\right)\cdot
(1+\gamma\cdot \ap(r)),\no &&
g(r)=\f{r^2}{L^2}\left(1-\f{r^6_0}{r^6}\right)\cdot (1+\gamma\cdot
\beta(r)),\no &&
\ap(r)=-\f{2\cdot 73}{5L^6}\left(67\f{r^6_0}{r^6}+67\f{r^{12}_0}{r^{12}}-37\f{r^{18}_0}{r^{18}}\right),\no
&&
\beta(r)=-\f{2\cdot 73}{5L^6}\left(67\f{r^6_0}{r^6}+67\f{r^{12}_0}{r^{12}}-245\f{r^{18}_0}{r^{18}}\right).
\ea
The AdS radius $L$ is related to the number $N$ of M2-branes by
\be L^9=N^{3}\f{4\kappa^2_{11}}{\pi^5}, \label{M5L} \ee
and the volume of $S^4$
is given by \be \mbox{Vol}(S^4)=\f{8\pi^2}{3}(L/2)^4. \ee The radius
in the $\theta$-direction is found to be
\be
2\pi R=\f{2\pi L^2}{3r_0}-\f{2^2\cdot 7\cdot 73\pi\gamma}{15 L^4r_0}.
\label{M5R}
\ee

Using \eqref{M5L},
the expansion parameter $\gamma$ \eqref{Mgamma} is reexpressed as
\be
\gamma=\f{2^{2/3}\cdot\pi^{16/3}}{3}\cdot \f{L^6}{N^2}.
\ee
It is useful to note
\be
\f{\mbox{Vol}(S^4)}{\kappa^2_{11}}=\f{2}{3\pi^3}\cdot\f{N^{3}}{L^5}.
\ee

In the presence of the deficit angle $\delta=2\pi(1-n)$, the higher derivative term $W_7$ behaves like
\be
W_7
=\f{2^2\cdot 3^2\cdot 5\cdot 73\cdot r_0^{24}}{L^8r^{24}}
+2\pi(1-n)\delta(\gamma_A)\cdot\f{2^3\cdot 3\cdot 73\cdot r^{18}_0}{5L^6 r^{18}}
+O(\gamma) +O\((1-n)^2\),
\ee
using the uncorrected AdS$_7$ soliton metric.
This leads to the contribution
\ba
S^{(W)}_A
&=&
\f{2^4\cdot 3\cdot 73\pi}{5L^6}
\cdot\f{r_0^{4}}{14}
\cdot\f{2\pi RV_3\cdot\mbox{Vol}(S^4)}{2\kappa_{11}^2}\cdot\gamma, \no
&=& \f{73}{35}\cdot \f{2^{14/3}\cdot \pi^{13/3}}{3^5}\cdot \f{NV_3}{R^3}
+ O(N^{-1}).
\ea
The contribution from the Einstein-Hilbert action can be found by applying the area formula to the corrected metric, as
\ba
S^{(EH)}_A &=& \f{4\pi^2 RV_3\cdot\mbox{Vol}(S^4)}{\kappa^2_{11}}\cdot \int^{r_{\infty}}_{r_0}dr \f{r^3}{L^3}\left(1-\f{2^4\cdot 13\cdot 73\cdot \gamma}{5\cdot L^6}\cdot\f{r^{18}_0}{r^{18}}\right),\no
&=& \f{8N^3V_3R}{3\pi L^8}\cdot \left(\f{1}{4}r^4_{\infty}
-\f{1}{4}r^4_0-\f{2^4\cdot 13\cdot 73}{5\cdot 14}\cdot\f{\gamma}{L^6}r_0^4\right) \no
&=& \f{2N^3V_3R}{3\pi L^8}r^4_{\infty}
- \f{2}{3^5\pi}\f{N^3V_3}{R^3}
- \f{2^{14/3}\cdot 73\cdot \pi^{13/3}}{3^5\cdot 35}\cdot\f{NV_3}{R^3}
+ O(N^{-1}).
\ea
Finally, the total sum of the finite term looks like
\ba
\left(S_A^{(EH)}+S_A^{(W)}\right)_{finite}&=&-\f{2}{3^5\pi}\cdot \f{N^3V_3}{R^3}
+ O(N^{-1}),
\ea
that is, the correction to $S_A$ vanishes in the leading order ($\sim N^1$).

\section{Holographic Entanglement Entropy in Gauss-Bonnet Gravities}
\label{sec:gb}

In this section, we will investigate holographic entanglement entropy
for the half space of the boundary of AdS soliton geometries
in Gauss-Bonnet gravities.

\subsection{4D Gauss-Bonnet Gravity}
We consider 4D AdS-Einstein-Gauss-Bonnet gravity,%
\footnote{\corrected{
Precisely speaking, some surface terms should also be included in the action
(except for a particular value of $\eta=1/2$ in the 4D case \cite{4DST})
and they would affect the divergent part of the entanglement entropy.
Since we focus on the finite part of it in this section,
we simply omit them here.
}}
\be
S = -\f{1}{16\pi G_N^{(4)}}
\int d^{4}x\s{g}\left[
-2\Lambda  + \cR + \f{\eta L^2}{2}\cL_{GB}
\right],
\ee
where
\be
\Lambda = -\f{3}{L^2},
\qquad
\cL_{GB} =
\cR_{\mu\nu\rho\sigma}\cR^{\mu\nu\rho\sigma} - 4\cR_{\mu\nu}\cR^{\mu\nu} + \cR^2.
\label{gbterm}
\ee

The Euclidean AdS soliton metric (= Euclidean AdS-Schwarzchild metric) is
\be
ds^2 = \f{L^2}{r^2f(r)}dr^2 + \f{r^2f(r)}{L^2}d\theta^2
+ \f{r^2}{L^2}(dt_{E}^2 + dx^2),
\ee
where
\ba
f(r) = 1-\left(\f{r}{r_0}\right)^3,
\qquad
\theta\sim\theta + 2\pi R,
\qquad
R = \f{2 L^2}{3r_0}.
\ea
The solution does not depend on the Gauss-Bonnet coupling $\eta$,
since the Gauss-Bonnet term is topological in 4D.

We define the subsystem A of the boundary time slice $t_{E}=0$ by $x>0$.
To compute the entanglement entropy $S_A$ holographically,
we introduce a conical deficit on the $(t_E,x)$-plane along with
the prescription of the replica method,
and the bulk curvature tensors at that time is again given by
the formula \eqref{curv}.
Then the contributions of the Einstein-Hilbert term and Gauss-Bonnet term
are computed as, respectively,
\ba
S_A^{(EH)}
&=& -\f{\pd}{\pd n}\log\f{Z^{(EH)}_n}{\big(Z^{(EH)}_1\big)^n}\bigg|_{n=1} \no
&=& \f{\mathrm{Area}(\gamma_A)}{4G_N^{(4)}} \no
&=& \f{2\pi R}{4G_N^{(4)}}\int_{r_0}^{r_\infty}dr \no
&=& \f{\pi L^2}{3G_N^{(4)}}\f{r_\infty}{r_0} - \f{\pi L^2}{3G_N^{(4)}},
\ea
and
\ba
S_A^{(GB)}
&=& -\f{\pd}{\pd n}\log\f{Z^{(GB)}_n}{\big(Z^{(GB)}_1\big)^n}\bigg|_{n=1} \no
&=& -\f{\eta L^2}{32\pi G_N^{(4)}}2\pi R \int_{r_0}^{r_\infty}dr\s{g}\f{16\pi f(r)}{r^2} \no
&=& -\f{\pi\eta L^2}{3G_N^{(4)}}
\left[\f{2r}{r_0} + \Big(\f{r_0}{r}\Big)^2 \right]_{r_0}^{r_\infty} \no
&=& -\f{2\pi L^2}{3G_N^{(4)}}\frac{r_\infty}{r_0}\eta + \frac{\pi L^2}{G_N^{(4)}}\eta
\quad (r_\infty\to\infty).
\ea
Therefore the total expression of $S_A$ is,
\be
S_A = (\mathrm{div}.) - \f{\pi L^2}{3G_N^{(4)}} + \frac{\pi L^2}{G_N^{(4)}}\eta.
\ee

\subsection{5D Gauss-Bonnet Gravity}
Next we consider 5D AdS-Einstein-Gauss-Bonnet gravity,
\be
S = -\f{1}{16\pi G_N^{(5)}}
\int d^{5}x\s{g}\left[
-2\Lambda  + \cR + \f{\eta L^2}{2}\cL_{GB}
\right],
\label{5dgb}
\ee
where
$\Lambda = -\f{6}{L^2}$
and the expression of $\cL_{GB}$ is the same as \eqref{gbterm}.
The value of the Gauss-Bonnet coupling $\eta$ is restricted to the range of
\be
-\f{7}{36} < \eta < \f{9}{100},
\label{etarange}
\ee
from the demand of causality in the Lorentzian version of this theory \cite{BM}.

The Euclidean AdS soliton solution in this theory is given by the metric
\be
ds^2 = \f{L^2}{r^2f(r)}dr^2 + \f{r^2f(r)}{L^2}d\theta^2
+ \f{r^2}{L_{AdS}^2}(dt_{E}^2 + dx^2 + dy^2),
\ee
where
\ba
f(r) = \f{1}{2\eta}\left(1 - \s{1-4\eta\left\{1-\Big(\f{r_0}{r}\Big)^4\right\}}\right),
&&\quad
f_\infty
= \lim_{r\to\infty}f(r)
= \f{2}{1+\s{1-4\eta}},
\no
L_{AdS}
= \f{L}{\s{f_\infty}},
&&\quad
\theta\sim\theta + 2\pi R,
\qquad
R = \f{L^2}{2r_0}.
\ea
It is useful to note that $f'(r_0) = \sfrac{4}{r_0}$ and
$f(r_0) = \lim_{r\to\infty}f'(r)=0$.
This spacetime is asymptotically an AdS$_5$ with the radius $L_{AdS}$.
This solution was found in \cite{MS}
in the spherical form, as a black hole solution.%
\footnote{\corrected{
Before it, asymptotically flat black hole solution for $\Lambda=0$
had been found in \cite{BD}.
The planar form of the AdS one we are using here was presented
in \cite{Cai} and discussed as an AdS soliton in \cite{CKW}.
}}

We define the subsystem $A$ as $x>0$ again,
and call the coordinate length along the $y$-direction $V_y$
(we take $V_y\to\infty$).
In this case,
\ba
S_A^{(EH)}
&=& \f{2\pi R\cdot V_y}{4G_N^{(5)}}\int_{r_0}^{r_\infty}\f{r}{L_{AdS}}dr \no
&=& \f{\pi V_y Rr_\infty^2}{4G_N^{(5)}L}\s{f_\infty} - \f{\pi V_y L^3}{16G_N^{(5)}R}\s{f_\infty}
\ea
and
\ba
S_A^{(GB)}
&=& -\f{2\pi R\cdot V_y}{4 G_N^{(5)}}\cdot\f{\eta}{L_{AdS}}
 \int^{r_\infty}_{r_0}dr\left[6f(r) + 6rf'(r) + r^2f''(r)\right]r \no
&=& -\f{\pi R V_y \eta}{2 G_N^{(5)}L_{AdS}}
 \left[3r^2f(r) + r^3f'(r)\right]^{r_{\infty}}_{r_0} \no
&=& -\f{3\pi V_yR r_\infty^2}{2G_N^{(5)}L}\cdot\eta f_\infty^{\sfrac{3}{2}}
 + \f{\pi V_yL^3}{2G_N^{(5)}R}\cdot\eta \s{f_\infty}.
\ea
The total expression of $S_A$ is,
\ba
S_A = S_A^{(EH)} + S_A^{(GB)}
&=&  \f{\pi V_yR r_\infty^2}{4G_N^{(5)}L}\cdot (1-6\eta f_\infty)\s{f_\infty}
 - \f{\pi V_yL^3}{16G_N^{(5)}R}\cdot (1-8\eta)\s{f_\infty}.
\label{5dgb_straight_sa}
\ea
One may find that the divergent part looks proportional to
the central charge 
\be
 a = \f{\pi L_{AdS}^3}{8G_N^{(5)}}(1- 6\eta f_\infty ),
\ee
by an appropriate way of parameter fixing.%
\footnote{
Fixing $L_{AdS}$, $R$, $V_y$, and 
$R_\infty\equiv\sfrac{Rr_\infty}{L_{AdS}}$ realizes it,
where $R_\infty$ is the ``physical radius'' of the $\theta$-circle on the boundary.
}
But we do not know whether it is sensible because the divergent part
depends on the way of regularization.

\section{Entropic Phase Transitions in 5D Gauss-Bonnet Gravity}
\label{sec:gbpt}

Here we would like to calculate the holographic entanglement entropy (HEE) for a subsystem $A$ which is defined by
a strip with the width $l$ in Gauss-Bonnet theories. We are interested in a discontinuous transition on the behavior of the holographic entanglement entropy for the AdS soliton
as we change the size $l$. In the absence of the higher derivatives, this has been
studied in \cite{NiTa,KKM,PP} and this phase transition is identified with the confinement/deconfinement transition. There is a critical value $l=l_c$ such that $\f{dS_A}{dl}=0$ (confined phase)
for $l>l_c$ and $\f{dS_A}{dl}>0$ (deconfined phase) for $l<l_c$. Qualitatively similar results has been recently reproduced in lattice gauge theory approaches \cite{La,Lb,Lc}. Below we would like to study how this behavior is
affected by higher derivative terms.

In the four dimensional Gauss-Bonnet gravity, the curvature
correction term in the HEE formula (\ref{hheein}) is topological and is determined only the Euler number of the
surface $\gamma_A$. Thus the $l$ dependence of its higher derivative correction gets trivial. Therefore below we will study the 5D Gauss-Bonnet gravity.

\subsection{Extremum equation for holographic entanglement entropy}

We employ the proposed generalization (\ref{hheein}) \cite{HMM,deBoer:2011wk} of the original HEE \cite{RT} in
the Gauss-Bonnet gravity. More precisely, including the boundary term, the HEE is
given by minimizing the functional
\be
S_A=\f{1}{4G_N^{(5)}}\int_{\gamma_A}dx^3\s{h}(1+\eta L^2 \cR)+\f{\eta L^2}{2G_N^{(5)}}\int_{\de\gamma_A}dx^2\s{h_b}\cK,  \label{hhee}
\ee
where $h$ is the induced metric on the three dimensional surface $\gamma_A$ which satisfies $\de \gamma_A=\de A$ and $\cR$ is the intrinsic curvature of $h$; $h_b$ is the induced metric on $\de\gamma_A$ and
$\cK$ is the trace of its extrinsic curvature. If we define a unit normal vector of the embedding of
$\de\gamma_A$ in $\gamma_A$ by $n^a$, then we have
\be
\int_{\de\gamma_A}dx^2\s{h_b}\cdot \cK=\int_{\de\gamma_A}dx^2 \de_{a}(\s{h_b}\cdot n^a)|_{r=r_\infty}.
\ee

We assume that the subsystem $A$ sits on the time slice $t=0$ and extends in $y$- and $\theta$-directions.
Therefore we can specify the profile of $\gamma_A$ by the embedding function
\be
x=x(r).
\ee
Its holographic dual surface $\gamma_A$ is given by a codimension three surface defined by
\be
t=0,\ \ x=x(r),\ \ -\infty<y<\infty, \ \ 0\leq \theta \leq 2\pi R.
\ee
The regularized length in $y$ is denoted by $V_y$ below.
The surface $\gamma_A$ is determined by extremizing
the functional (\ref{hhee}). When there are several extremal surfaces, we select the one which has the smallest
value of (\ref{hhee}). We choose the boundary condition of $\gamma_A$ such that it starts from $x=\f{l}{2}$ at $r=r_\infty$, extends into the smaller $r$ region until it reaches $r=r_{*}$, and comes back to the AdS boundary
$r=r_{\infty}$ at $x=-\f{l}{2}$.

The induced metric on $\gamma_A$ is given by
\be
ds^2=h_{ab}dx^adx^b=\left(\f{L^2}{r^2f(r)}+\f{r^2}{L^2_{AdS}}(x'(r))^2\right)dr^2+\f{r^2f(r)}{L^2}d\theta^2
+\f{r^2}{L^2_{AdS}}dy^2.
\ee
We obtain
\ba
&& \s{h}=\f{r}{L\cdot L^2_{AdS}}\s{L^2\cdot L^2_{AdS}+r^4f(r)x'^2},\no
&& \s{h}\cR=\f{2rf(r)+r^2f'(r)}{L\s{L^2\cdot L^2_{AdS}+r^4f(r)x'^2}}+q'(r), \label{eem}
\ea
where we defined
\be
q(r)=-\f{r^3f'(r)+4r^2f(r)}{L\s{L^2\cdot L^2_{AdS}+r^4f(r)x'^2}}.
\ee
When we integrate on $\gamma_A$, the second term $q'(r)$ in (\ref{eem}) yields surface terms. We can show that the contribution of $q(r_{\infty})$
is canceled by the Gibbons-Hawking term in (\ref{hhee}).
Furthermore,
since the profile is connected with its mirror image at $r=r_*$,
there is indeed no boundary there and neither is the boundary term $q(r_*)$.
Therefore the functional we need to minimize is given by
\ba
S_A&=&\f{\pi RV_y}{G_N^{(5)}}
\int^{r_{\infty}}_{r_*}\f{L^2\cdot L_{AdS}^2\cdot r(1+g(r))+r^5f(r)x'^2}{L\cdot L^2_{AdS}\s{L^2\cdot L^2_{AdS}+r^4f(r)x'^2}}dr
\no
&=&\f{\pi RV_y}{G_N^{(5)}\cdot L_{AdS}}
\int^{r_{\infty}}_{r_*}\f{rg(r)+rX}{\s{X}}dr,
\label{heeo}
\ea
where
\ba
g(r)&=&2\eta f(r)+\eta rf'(r),\\
X &=& 1 + \f{1}{L^2L_{AdS}^2} r^4f(r)x'^2,
\ea
and $r=r_*$ is the turning point of the surface.
Note that $X\ge 1$.

The equation of motion for (\ref{heeo}) leads to the conservation law
\be
\f{r^5f(r)x'(L^2\cdot L^2_{AdS}+r^4f(r)x'^2-L^2L_{AdS}^2\,g(r))}{\left(L^2\cdot L^2_{AdS}+r^4f(r)x'^2\right)^{3/2}}=r_s^3\s{f(r_s)},
\label{5dgbcl}
\ee
where $r_s$ is some constant.
If the surface is smooth at $r=r_*$, i.e.,
$x'(r_*)$ gets divergent,
then $r_s=r_*$.
We can find the width of the subsystem $A$ from
\be
l=2\int^{r_{\infty}}_{r_*}dr~ x'(r).
\label{5dgbl}
\ee

Now let us examine the equation \eqref{5dgbcl}.
We impose the obvious inequality:
\be
f_\infty > 0 \quad \Leftrightarrow \quad \eta \le \f{1}{4}.
\ee
Although the causality bounds for $\eta$ is known to be given by \eqref{etarange} \cite{BM},
at first we would like to proceed by ignoring it and would like to come back to this in the end.
By taking the square of \eqref{5dgbcl}, we obtain
\be
\f{F(X)}{X^3} = 0, \label{eomq}
\ee
where
\ba
F(X) &=& (r^6f(r)-r_s^6f(r_s))X^3 - r^6f(r)(1+2g(r))X^2 \no
&&+ r^6f(r)(2g(r)+g(r)^2)X - r^6f(r)g(r)^2.
\label{FX}
\ea

\subsection{Various phases for ``minimal surfaces''}
Let us examine the solutions of the equation (\ref{5dgbcl}).
In order to have a physical solution for the surface $\gamma_A$, we need
a solution to the cubic equation (\ref{eomq}) with $X(r)\geq 1$. This constrains
the range of $r$ where desirable solutions exist.

\begin{figure}
\centering
\includegraphics[width=0.3\textheight,origin=c,angle=-90]{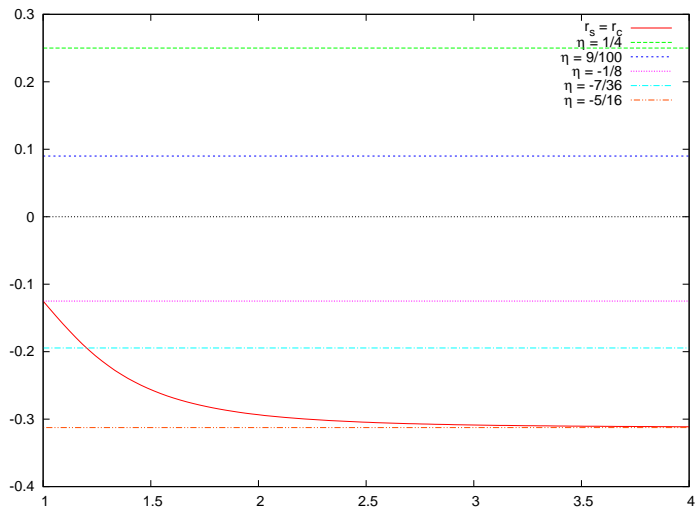}
\vspace{-1.5cm}
\caption{The bounds for $g(r_s)\ge -1/2$ \eqref{paramrestriction01}\eqref{rc}
(red) on the $(r_s/r_0, \eta)$-plane.
Lines for $\eta = 1/4, 9/100, -1/8, -7/36, -5/16$ are also displayed.}
\label{fig:smoothcond}
\end{figure}

The existence of smooth solutions to (\ref{eomq}) requires that $X$ is positively divergent
at the turning point $r=r_*$. In this case we find $r_*=r_s$ as follows from (\ref{5dgbcl}).
This requires that a solution to (\ref{eomq}) gets positively divergent as we approach
$r=r_s$ from the above. This condition leads to the inequality $g(r_s) \ge -\sfrac{1}{2}$ or equally,
\be
\eta \ge -\f{1}{32}\f{{\tilde{r}_s}^4 - 9 + 3\sqrt{9{\tilde{r}_s}^8 - 2{\tilde{r}_s}^4 + 9}}{{\tilde{r}_s}^4},
\label{paramrestriction01}
\ee
where $\tilde{r}_s \equiv r_s/r_0$. This is always satisfied when
\be
\eta \ge -\f{1}{8}.
\ee
Also, for the values
\be
-\f{5}{16} < \eta < -\f{1}{8},
\ee
only a particular range of $r_s$
\be
r_s \ge r_c\equiv
\f{\s{3}}{2}\(\f{-\eta}{(\f{1}{4}-\eta)(\f{5}{16}+\eta)}\)^{\f{1}{4}}r_0,
\label{rc}
\ee
is allowed. For $\eta<-\f{5}{16} $ there will be no smooth solutions. See Fig.\ref{fig:smoothcond} for
numerical plots of the bound.

\begin{figure}
\centering
\begin{tabular}{ccc}
\includegraphics[width=0.3\textwidth,origin=c,angle=180]{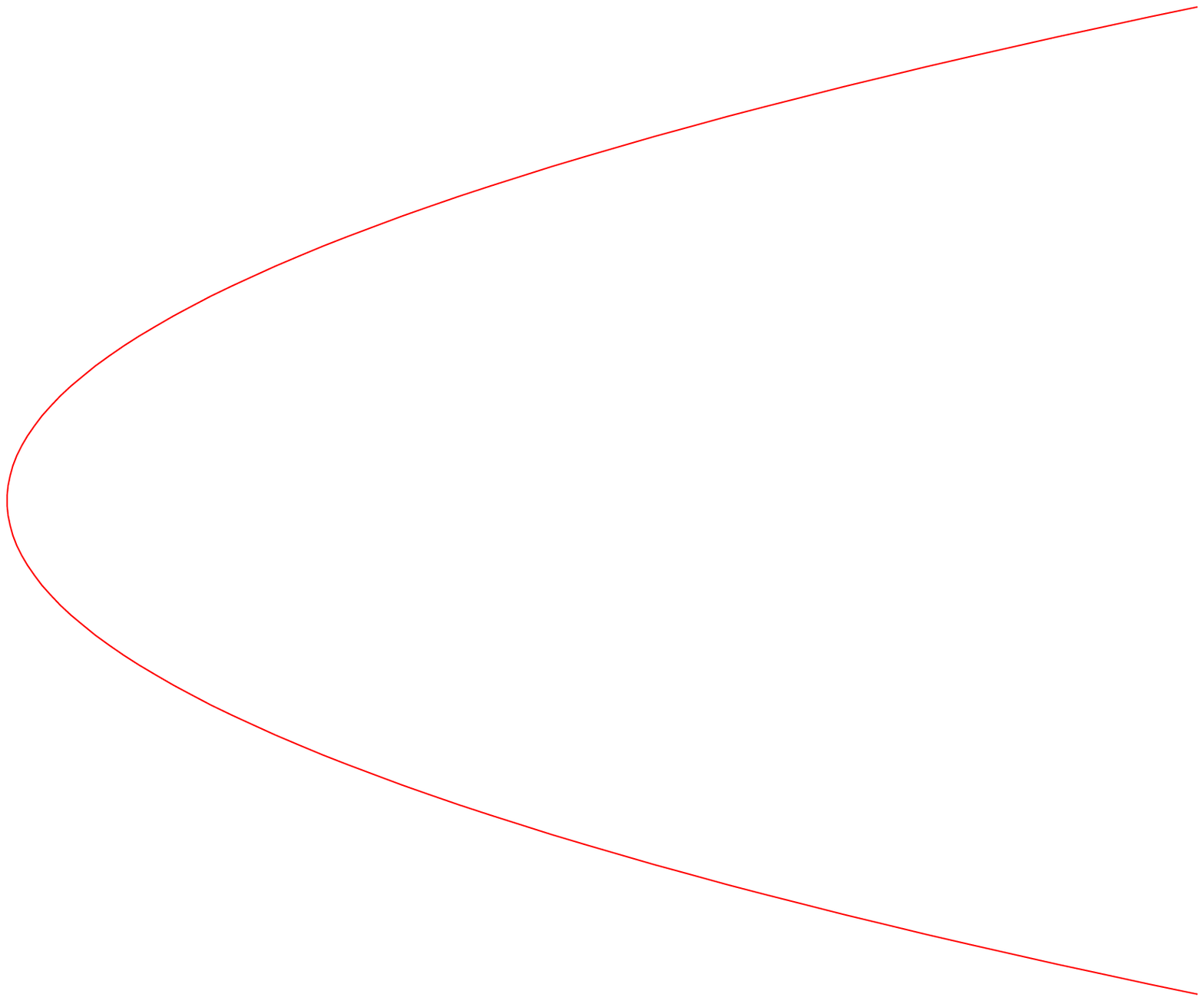}
&
\includegraphics[width=0.3\textwidth,origin=c,angle=180]{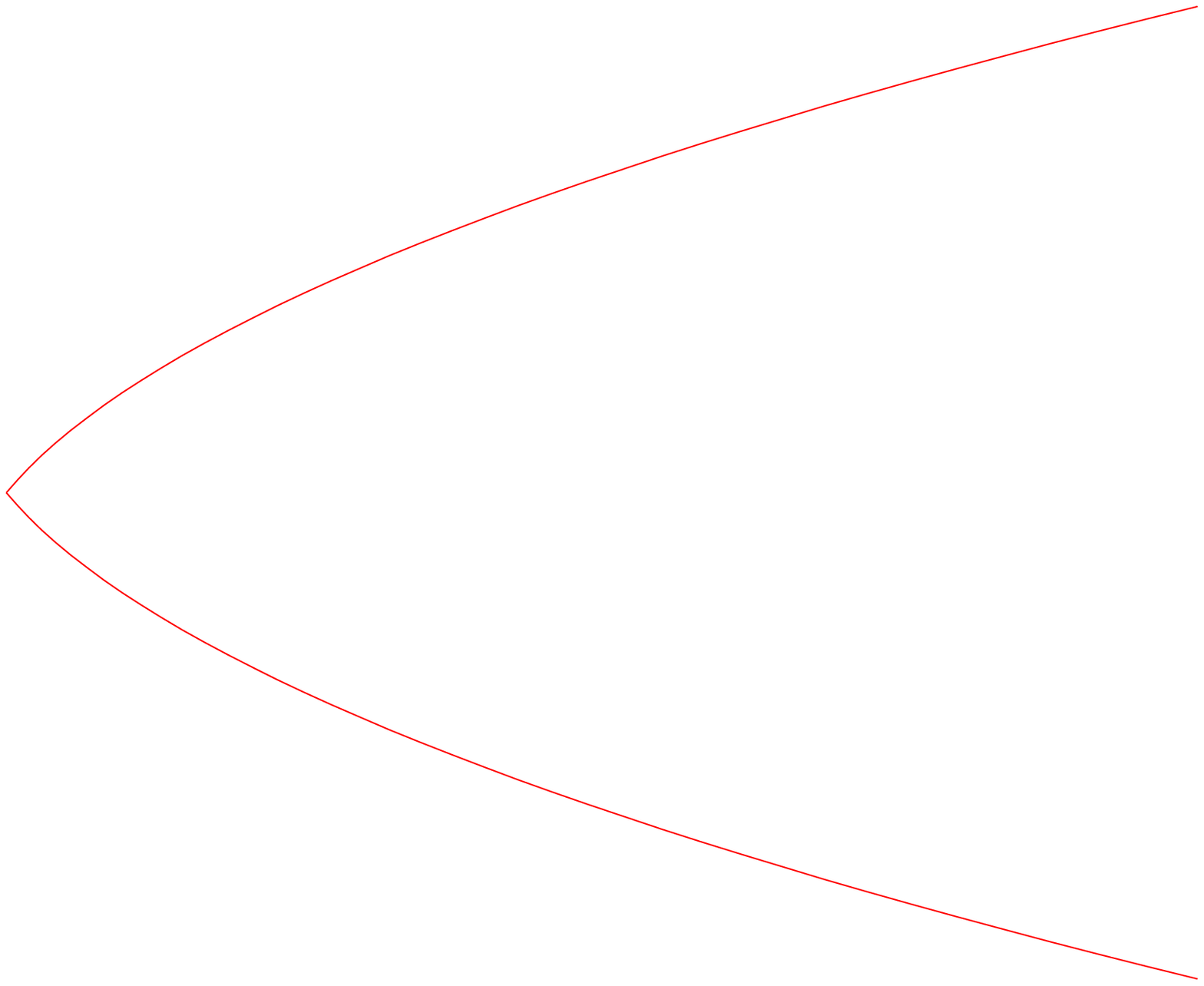}
&
\includegraphics[width=0.3\textwidth,origin=c,angle=180]{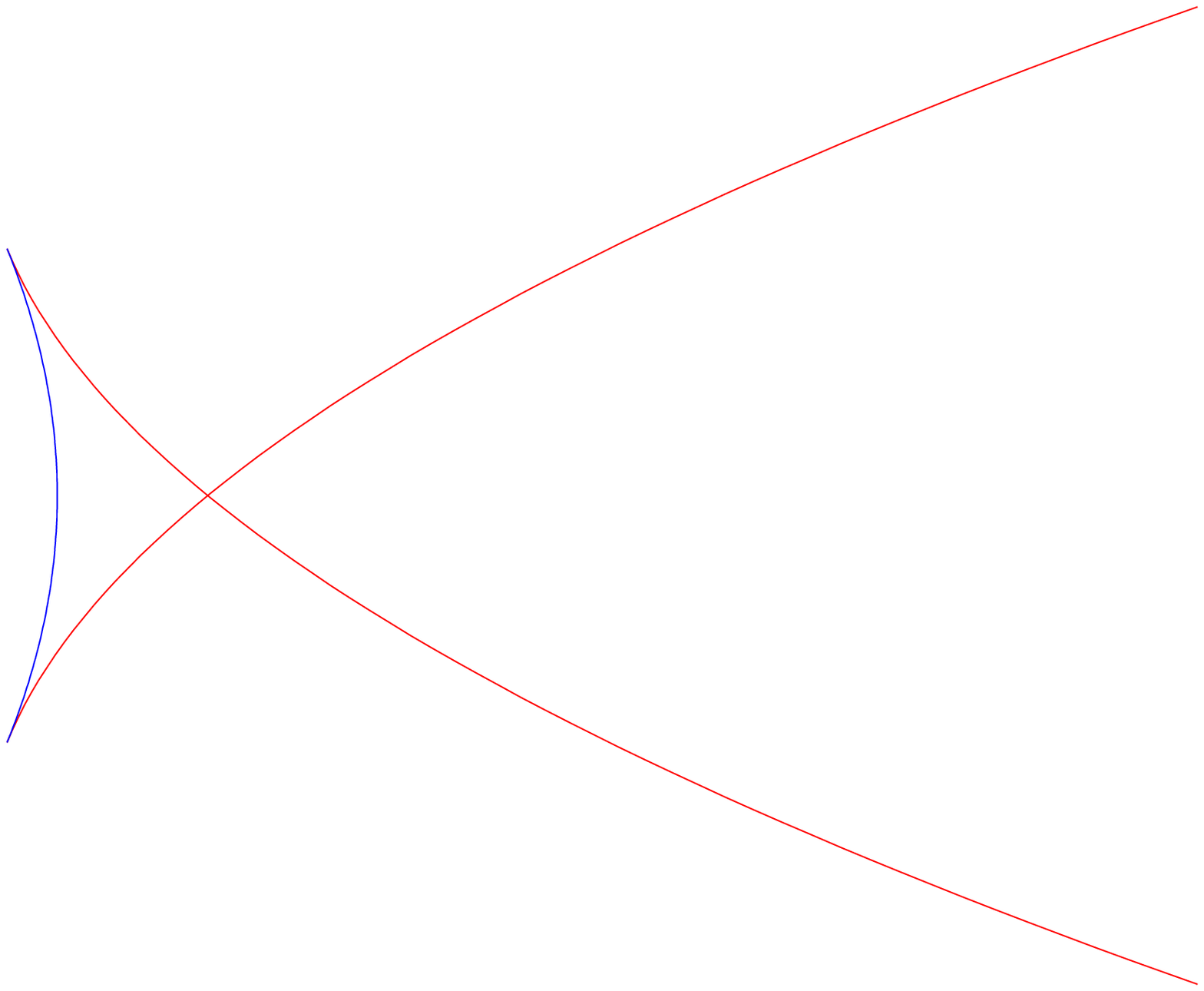}
\\
(a) smooth surface
&
(b) surface singular at $r = r_*$
&
(c) ``switch-back'' surface
\end{tabular}
\caption{
Patterns of extremal curved surfaces $\gamma_A$ (projected on the $(r,x)$-plane).
The integrand of \eqref{heeo} is locally minimum on the red lines
and maximum on the blue one.
Therefore we only have to consider (a) and (b).
}
\label{fig:profiles1}
\end{figure}

In these cases, a solution to (\ref{eomq}) satisfies $X> 1$ for $r>r_s$
and it diverges at $r=r_s$ while
there is no such solution for $r<r_s$. Moreover by studying the second order perturbations,
it is indeed a local minimum of the functional (\ref{hhee}). In this way, we find a smooth solution with the correct boundary condition and thus this gives the first candidate for $\gamma_A$ (as depicted in Fig.\ref{fig:profiles1}(a)). 

Now we would like to turn to the other cases with $g(r_s) < -1/2$. Since we cannot 
find smooth solutions, we need to be satisfied with a solution which is not smooth at the turning point $r=r_*$ as e.g. the one depicted in Fig.\ref{fig:profiles1}(b), assuming that 
the minimal surface always exists\footnote{In the previous case with $g(r_s) \geq -1/2$,
we can also construct a similar non-smooth surface like Fig.\ref{fig:profiles1}(b). However, 
we can confirm that the functional (\ref{hhee}) is always greater than the smooth one.}. The existence of solution leads to another bound $r>r_b$, where $r_b$ satisfies
\be
r_s^6 f(r_s) = \f{4}{27}\f{(g(r_b)-1)^3}{g(r_b)}\cdot r_b^6f(r_b).
\ee
For $r>r_b$, there are two solutions which satisfy $X>1$.
The smaller is continuous through $r=r_s(>r_b)$, and the greater is
divergent in the limit $r\to r_s - 0$.
They get coincident at $r=r_b$ and disappear for $r<r_b$.
Joining these, we can consider, for example, a ``switch-back''
surface like Fig.\ref{fig:profiles1}(c).
However, the second order perturbation of \eqref{heeo}
is positive for the smaller solution, and negative for the greater.
Then we need to consider only the first,
and in this case the form of the surface is always like Fig.\ref{fig:profiles1}(b).

\begin{figure}
\centering
\begin{tabular}{cc}
\includegraphics[height=0.4\textwidth,origin=c,angle=-90]{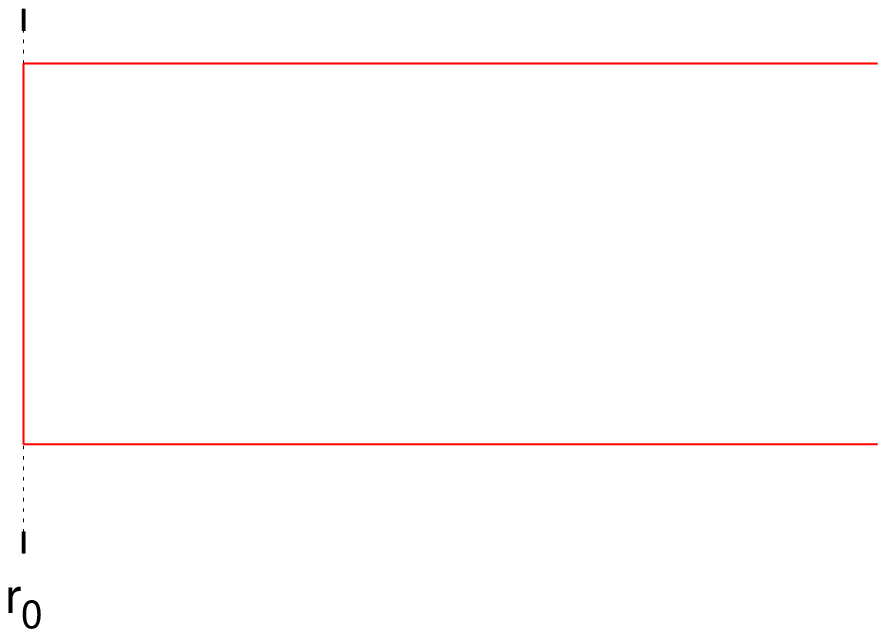}
\vspace{-0.7cm}
&
\includegraphics[height=0.4\textwidth,origin=c,angle=-90]{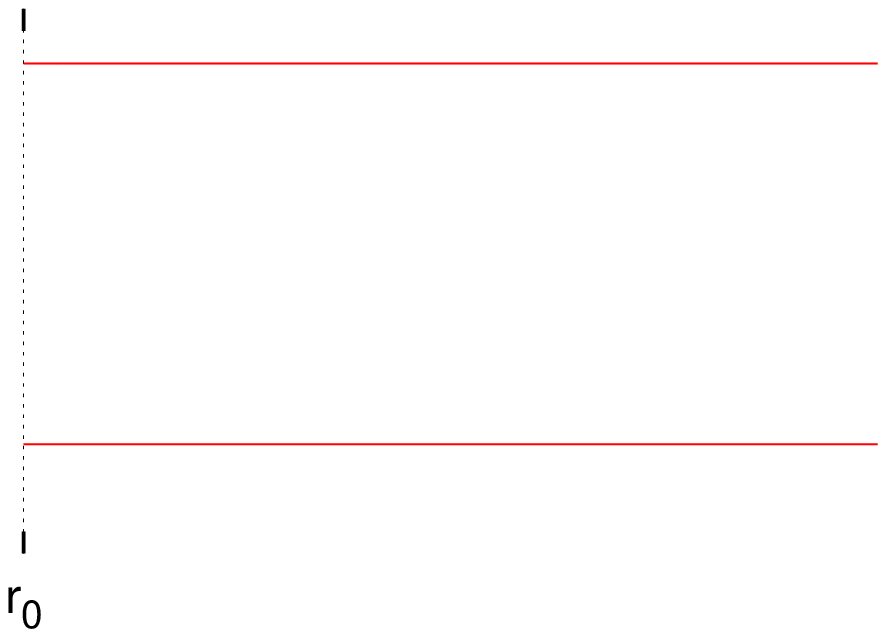}
\vspace{-0.7cm}
\\
(d) connected at $r=r_0$
&
(e) disconnected
\end{tabular}
\caption{
Two patterns of ``trivial'' $\gamma_A$.
They have different values of $S_A$ due to the boundary term $q(r_0)$,
accounted only for (e).
For $\eta < 0$, (e) is preferred to (d), and vice versa for $\eta > 0$.
}
\label{fig:profiles2}
\end{figure}
Other than the solutions discussed above,
there are two ``trivial'' candidates of the ``minimal surface''.
One is displayed in Fig.\ref{fig:profiles2}(d),
which is a pair of the surfaces $x=-l/2$ and $x=l/2$ connected at $r=r_0$.
It is not a solution of the extremum equation $F(X)=0$
but on the edge of the configuration space of $x(r)$.
The corresponding value of $S_A$ is,
independently from $l$, given by substituting
$r_*=r_0$ and $X\equiv 1$ to \eqref{heeo}, resulting
\ba
S_A^{(d)}
&=& \f{\pi RV_y}{G_N^{(5)}\cdot L_{AdS}}\int_{r_0}^{r_\infty}(rg(r)+r)dr \no
&=& \f{\pi RV_y}{2G_N^{(5)}\cdot L_{AdS}}\left[\left(1 + 2\eta f_\infty\right) r_\infty^2  - r_0^2\right].
\label{sad}
\ea
The other, which is displayed in Fig.\ref{fig:profiles2}(e),
is similar to it but different in an important way.
Unlike the all variations of the surfaces discussed above,
it consists of two parts which are not connected to each other.
Therefore the corresponding value of $S_A$ includes the contribution from the boundary term $q(r_0)=-\sfrac{4r_0^2}{L^2L_{AdS}}$ and
\ba
S_A^{(e)}
&=& S_A^{(d)} - \f{2\pi R V_y\cdot \eta L^2}{4G_N^{(5)}}q(r_0) \no
&=& \f{\pi RV_y}{2G_N^{(5)}\cdot L_{AdS}}\left[\(1 + 2\eta f_\infty\) r_\infty^2  - \(1-8\eta\)r_0^2\right].
\label{sae}
\ea
This is, as is expected, precisely twice of \eqref{5dgb_straight_sa}
except the difference of the divergent part coming from the boundary term.%
\footnote{
In this section we take the last term in \eqref{hhee} into consideration,
while we omitted the boundary term in \eqref{5dgb}.
In fact, if we were to omit the boundary term in \eqref{hhee},
we have an additional contribution from $q(r_\infty)$ to \eqref{sae}
and the result agrees with twice of \eqref{5dgb_straight_sa} exactly.
}
We can immediately see that,
when $\eta<0$ the phase (e) is preferred to (d),
and for $\eta>0$ (d) is preferred.

\subsection{Numerical results and phase structures}

\begin{figure}
\centering
\includegraphics[width=0.4\textheight,angle=-90]{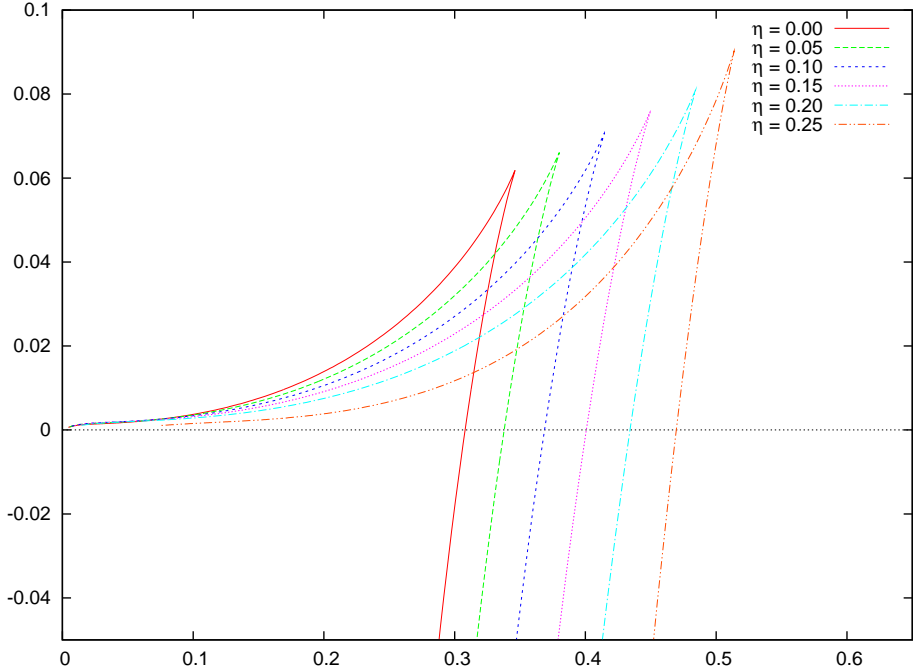}
\caption{$(l,S_A)$ plots for variable $0 \le \eta \le \sfrac{1}{4}$.
The unit for $l$ and $S_A$ are \corrected{$R$} and 
$\sfrac{\pi V_yL}{2G_N^{(5)}}$, respectively.
The value for the phase (d) is taken to be $0$ for $S_A$.}
\label{fig:5dgb_plots3}
\end{figure}

\begin{figure}
\centering
\includegraphics[width=0.4\textheight,angle=-90]{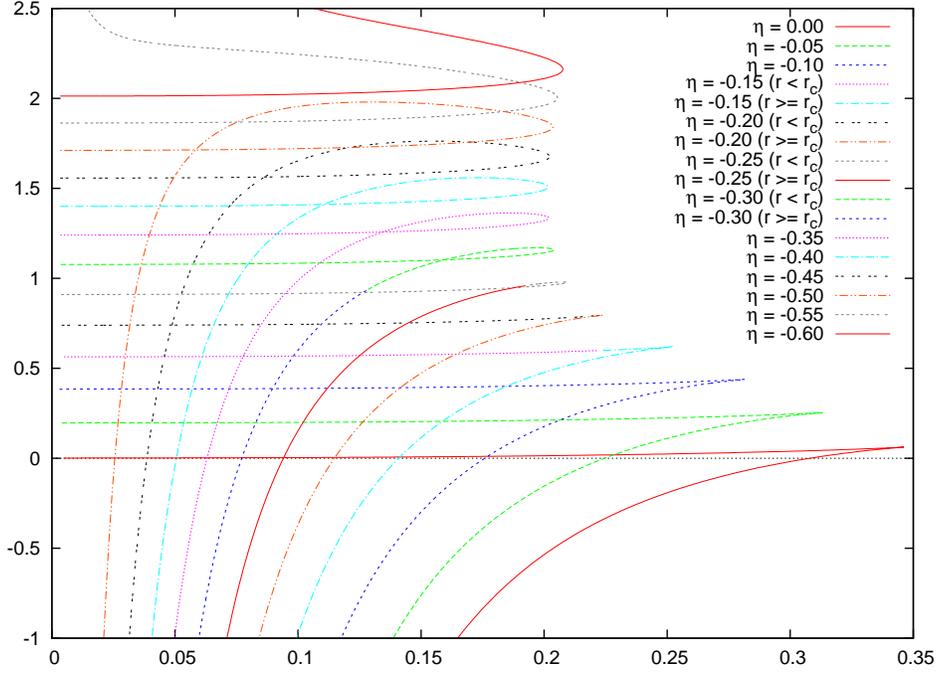}
\caption{$(l,S_A)$ plots for variable $\eta\le 0$.
The value for the disconnected phase (e) is taken to be $0$ for $S_A$.}
\label{fig:5dgb_plots}
\end{figure}

Now we can numerically execute the integrals \eqref{heeo} and \eqref{5dgbl}
for the solutions of the extremum equation \eqref{5dgbcl},
and plot the values on the $(l, S_A)$-plane.
We set the turning point $r_*$ as $r_*=r_s$ when $g(r_s)\geq -\f{1}{2}$, while 
we set it as $r_*=r_b$ when $g(r_s)< -\f{1}{2}$. These are because we find that 
they give the smallest
values of the functional (\ref{hhee}) when we change $r_*$ with $l$ fixed.
The resulting behavior of $S_A(l)$ is plotted in 
Fig.\ref{fig:5dgb_plots3} ($\eta\ge 0$)
and
Fig.\ref{fig:5dgb_plots} ($\eta\le 0$).
The zero point of $S_A$ is taken to be
$S_A^{(d)}$ \eqref{sad} for $\eta\ge 0$
and
$S_A^{(e)}$ \eqref{sae} for $\eta\le 0$.
The case for $\eta=0$ is the same one that was investigated in \cite{NiTa,KKM}.

Where the curve $S_A=S_A(l)$ is below the $l$-axis, the corresponding nontrivially 
connected surface, given as the solution of the extremum equation,
is the ``minimal surface'' $\gamma_A$  for each $l$.
Otherwise the minimal surface is given by the ``trivial'' surfaces
(d) or (e), depending on the sign of $\eta$.
Therefore, when the line crosses the $l$-axis, there occurs a phase transition
between those different phases as we mentioned before. 

First look at Fig.\ref{fig:5dgb_plots3}.
We can see that the qualitative form of the curve does not change in the range
$0\le\eta\le\sfrac{1}{4}$.
The phase of the curved surface is preferred when $l$ is small,
and
there occurs a phase transition at some particular value $l=l_c$,
for example, $l_c\corrected{/R}\simeq 0.338$ for $\eta=0.05$.
Notice that nothing special happens around the causality upper bound $\eta=0.09$ \eqref{etarange}.

\begin{figure}
\centering
\begin{tabular}{ccc}
\includegraphics[height=0.3\textwidth,angle=-90]{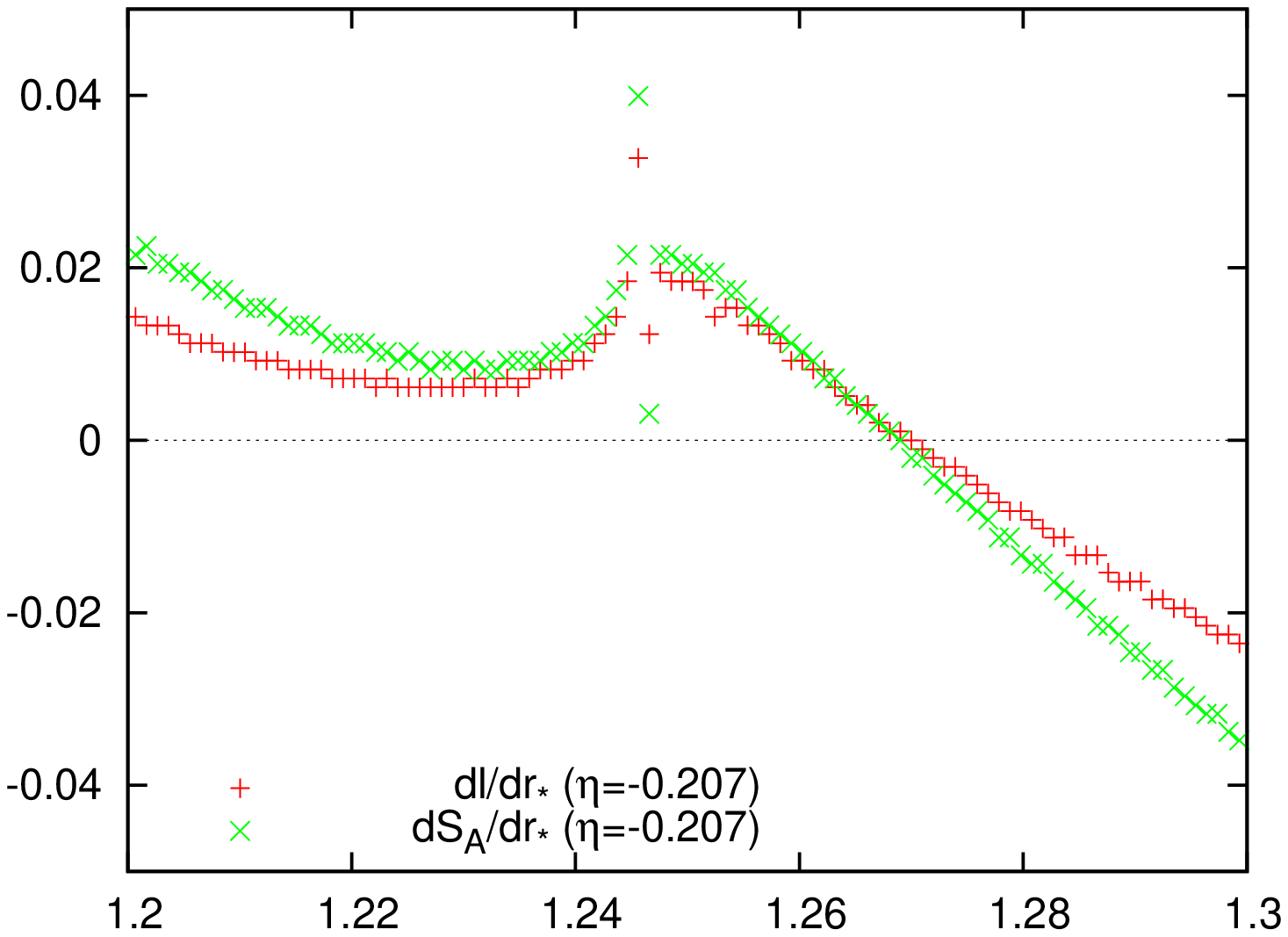}
&
\includegraphics[height=0.3\textwidth,angle=-90]{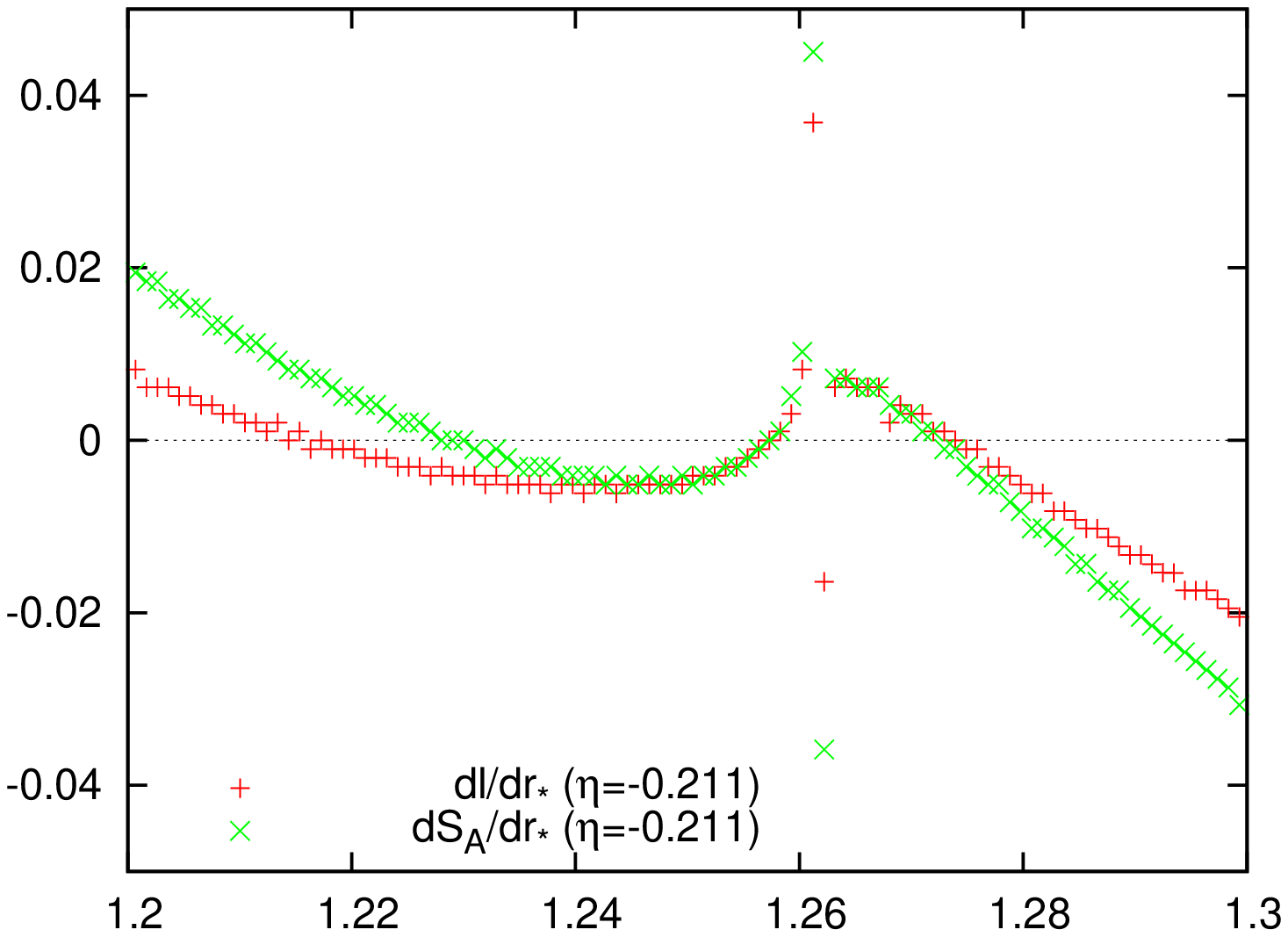}
&
\includegraphics[height=0.3\textwidth,angle=-90]{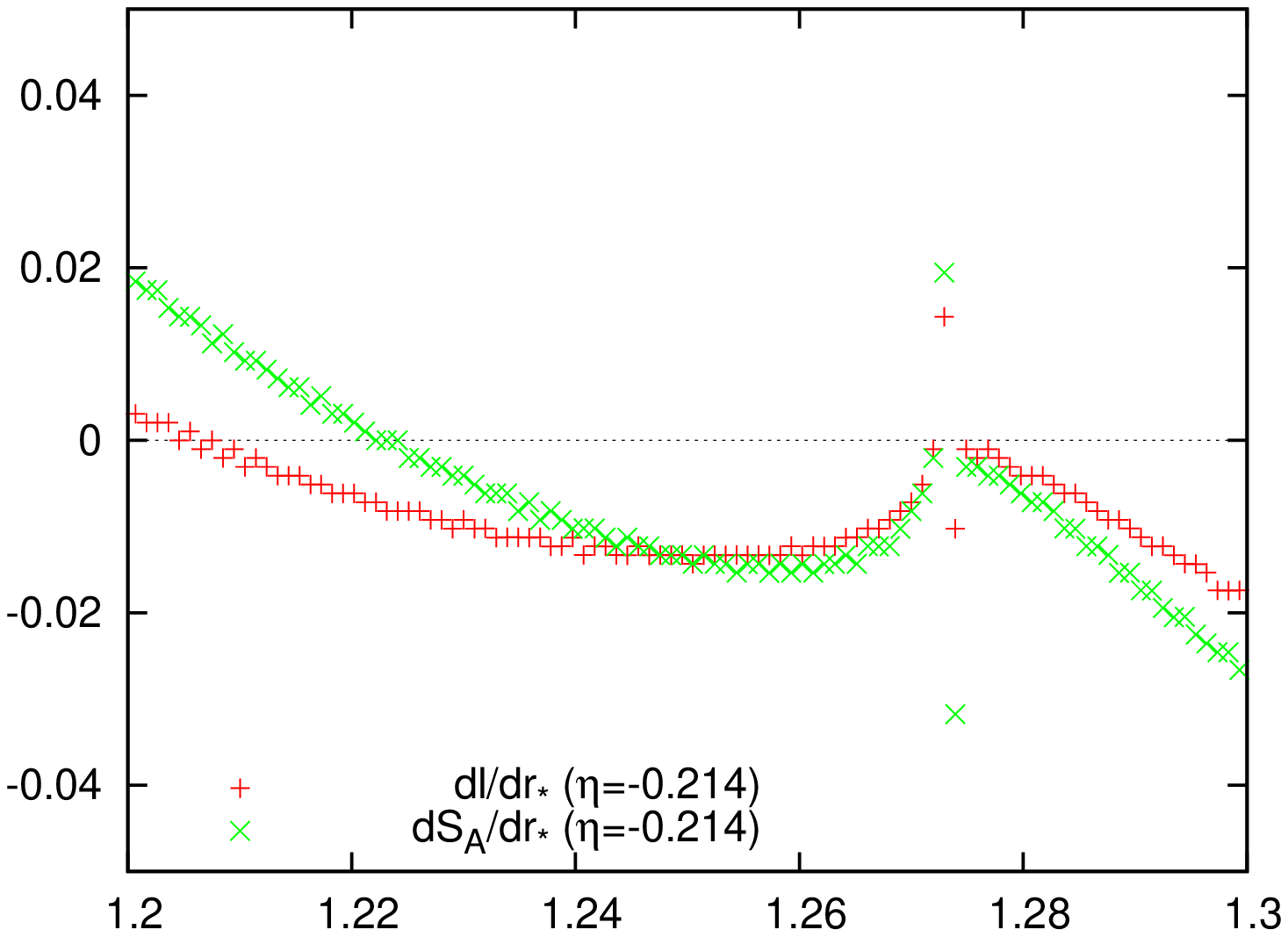}
\\
$\eta=-0.207$ ($r_c\simeq 1.2466$)
&
$\eta=-0.211$ ($r_c\simeq 1.2620$)
&
$\eta=-0.214$ ($r_c\simeq 1.2739$)
\end{tabular}
\caption{Plots for $dl/dr_*$ (red) and $dS_A/dr_*$ (green).}
\label{fig:loops}
\end{figure}

\begin{figure}
\centering
\includegraphics[width=0.4\textheight,angle=-90]{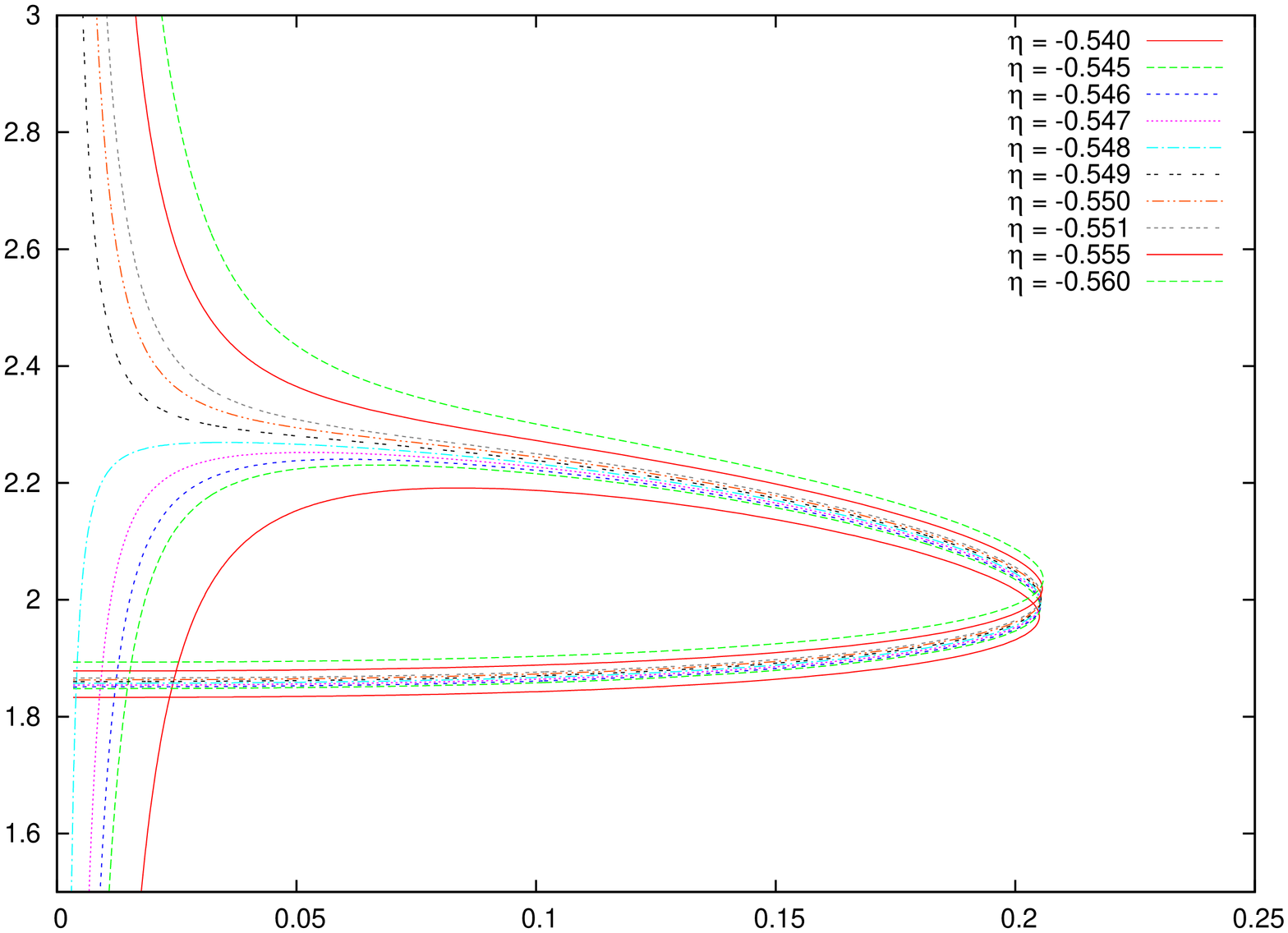}
\caption{$(l,S_A)$ plots for $\eta$'s around $-0.55$.}
\label{fig:5dgb_plots2}
\end{figure}

The plots for $\eta<0$, displayed in Fig.\ref{fig:5dgb_plots},
may be more interesting.
As we lower the value of $\eta$, the plotting curve changes its shape.
Below $\eta=-1/8=-0.125$, the point corresponding to the boundary
$r_*=r_c$ \eqref{rc} between (a) and (b)
walks on the curve from the origin toward the turning point.
Just before it reaches there, when $\eta\simeq-0.209$,
one (or more) loop(s) appears around there (Fig.\ref{fig:loops}).
After that $r_c$ continues to walk downward until $\eta=-5/16=-0.3125$.
Regardless the phenomena above, the phase structure is not altered.
However, across $\eta=-0.548\dots$,
the shape of the plotting curve changes dramatically, and
below there, the curve is above the $l$-axis everywhere and so
the disconnected phase is always favored (Fig. \ref{fig:5dgb_plots2}).
Note that these strange phenomena do not take places above the
causality lower bound $\eta=-7/36=-0.1944\dots$ \eqref{etarange}.

In summary, if we impose the causality bound \eqref{etarange}, none of new phenomena found for 
the higher derivative theory such as  the non-smooth surface as $\gamma_A$
like Fig.\ref{fig:profiles1}(b), the loopy profile of $S_A(l)$, or the absence of phase transition, do not occur. In this sense, we can conclude that the higher derivative corrections does not largely change qualitative properties of HEE for AdS solitons. If we temporally 
forget the causality bound, then from the above analysis we can learn that there should be lower bound for $\eta$ otherwise the strange behaviors start to occur. For example, if we 
require that the non-smooth surfaces should not appear as $\gamma_A$, then we obtain the 
bound $\eta>-5/16$. Also, if we exclude the absence of phase transition, then we find 
$\eta>-0.548\ddd$. Moreover, if we consider the Gauss-Bonnet gravity 
literally without considering more higher derivative terms, then we will find that HEE is not well-defined for the region $\eta>0$ as we will explain in the next subsection.

\subsection{Comments on an instability for $\eta>0$}

Before we conclude this paper, we would like to point out an important fact for the 
 conjectural formula \eqref{hhee} in the Gauss-Bonnet gravity.
Consider a smooth surface $\gamma_A$ which minimizes the functional (\ref{hhee}).
We can assume that it is symmetric along $\theta$. Let us focus on the region near the turning point $r=r_*$, where we can treat the warp factor of the AdS space as a constant.
In that region, we can simply ignore the $\theta$-direction and we can regard
$\gamma_A$ as an effectively 2-dimensional surface in a flat 
3-dimensional ambient space, spanned by $(r,x,y)$. In this setup, 
we can add infinitesimally small handles to $\gamma_A$ near $r=r_*$ and
decrease the value of the $\int\!\sqrt{h}\cR$ (i.e. the Euler number) without changing the other terms. Thus if we assume $\eta>0$,
we can take \eqref{hhee} to be an arbitrarily small value and so there is no minimum of the this functional. In this way we find that HEE $S_A$ is ill-defined for $\eta<0$ in the Gauss-Bonnet 
gravity. This argument gets more clear in 4D Gauss-Bonnet gravity as the curvature 
contribution in \eqref{hhee} gets purely topological without focusing on near the turning points. 

One may worry that our analysis for $\eta>0$ may be meaningless. 
However, this is not the case if
we implicitly assume the presence of more higher derivative corrections in addition to the Gauss-Bonnet gravity, which will be the case in string theory.  Indeed, we can show that the problem we mentioned occurs only when the higher order corrections are absent. In other words, what we find here is that if we consider the purely Gauss-Bonnet gravity in any dimension literally without any more higher derivative terms, then we will find that HEE is not well-defined when $\eta>0$.

\section{Discussions and Summary}
\label{sec:summary}

In this paper, we studied the holographic entanglement entropy (HEE) for AdS soliton geometries in the presence of higher derivative corrections. Our results in this paper show that the proposed higher derivative correction to HEE (\ref{hhee}) behaves in a sensible way.

In the first half part, we calculated the 
leading higher derivative corrections due to $\cR^4$ term for AdS soliton geometries in string theory and M-theory. Via the AdS/CFT, they are dual to the strong coupling expansions 
in the dual confining gauge theories such as a compactified $\cN=4$ super Yang-Mills. Our result is qualitatively consistent with the free Yang-Mills calculation of the entanglement entropy.

In the latter half part, we studied the HEE for AdS solitons in the Gauss-Bonnet gravity. 
Especially we examined the dependence of HEE on the size of the subsystem $A$. 
If we restrict to the known causality bound of the Gauss-Bonnet parameter $\eta$, our result shows that the behavior of HEE and the structure of phase transition are qualitatively similar to that for Einstein gravity i.e. $\eta=0$. However, if we
go beyond that bound, we observed several strange behavior such as the absence of phase transition and singular behavior of the solutions. We also find that if we consider the purely Gauss-Bonnet gravity literally without any more higher derivative terms, then we will find that HEE is not well-defined when $\eta>0$.

\section*{Acknowledgements}
The authors are grateful to
Jan de Boer,
Mitsutoshi Fujita, Shinji Hirano, Robert Myers, Masaki Shigemori, Shigeki Sugimoto and Tomonori Ugajin
for valuable discussions and comments. They are supported by World Premier International Research Center Initiative (WPI Initiative) from the Japan Ministry of Education, Culture, Sports, Science and Technology (MEXT).
N.O. is supported by the postdoctoral fellowship program of
the Japan Society for the Promotion of Science (JSPS),
and partly by JSPS Grant-in-Aid for JSPS Fellows No.\,22-4554. T.T. is very
grateful to the Aspen center for physics and the Aspen workshop
``Quantum Information in Quantum Gravity and Condensed Matter
Physics,'' where a part of this work was conducted.
T.T. is partly supported by JSPS Grant-in-Aid for Scientific Research No.\,20740132
and by JSPS Grant-in-Aid for Creative Scientific Research No.\,19GS0219.

\end{document}